\magnification = 1040
\baselineskip16.5pt
\abovedisplayskip 6pt plus2pt minus5pt
\belowdisplayskip 6pt plus2pt minus5pt
\vsize=9.7 true in \raggedbottom
\hsize=5.7 in
\voffset= -0.1 true in
\font\bmit=cmmib10 \textfont9=\bmit \def\bmit{\fam9 }
\font\bx=cmbsy10 \textfont10=\bx \def\bx{\fam10 }

\mathchardef\pi="7119
\mathchardef\sigma="711B
\mathchardef\mu="7116
\mathchardef\nabla="7272
\parskip=8pt plus3pt
\interlinepenalty=1500
\def\boxit#1{\vbox{\hrule\hbox{\vrule\kern1pc
   \vbox{\kern1pc#1\kern1pc}\kern1pc\vrule}\hrule}}
\def\sqr#1#2{{\vcenter{\vbox{\hrule height.#2pt
   \hbox{\vrule width.#2pt height#1pt  \kern#1pt
      \vrule width.#2pt}
     \hrule height.#2pt }}}}

\centerline{\bf THE POSITIVE RADIAL MOMENTUM OPERATOR}

\rightline{Shaun N Mosley,\footnote{${}^*$} {E-mail:
shaun.mosley@ntlworld.com \hfil\break
 correspondence: Sunnybank, Albert Rd, Nottingham NG3 4JD, UK}
Alumni, University of Nottingham, UK }

\vskip 0.3 in

\noindent {\bf Abstract } \quad
The Laplacian in spherical coordinates contains the squared radial
momentum operator
$ p_r^2 = - r^{-1} \partial_r^2 \, r \, ,$ which is Hermitian
and positive. However as has been pointed out by various authors the
``radial momentum operator"
$ \tilde{p}_r \equiv - i r^{-1} \partial_r r \, $ is not Hermitian.
The positive square root
operator $ p_r^+ \, $ such that $ ( p_r^+ )^2 = p_r^2 \, $ is found
and its inverse
$ | p_r^+ |^{-1} \, .$ We discuss the relation of these operators with
Fourier transforms, the Hilbert transform and fractional integral operators.

\vskip 0.1 in

\beginsection I. Introduction

Radial momentum operators naturally arise in quantum mechanics whenever
the Hamiltonian contains a radial potential term. In spherical
coordinates
$$ - \nabla^2  = - r^{-1} \partial_r^2 \, r  \, + \, L^2 / r^2 \, $$
where $ {\bf L} \, $ is the angular momentum operator. The squared
radial momentum operator
$$ p_r^2 \equiv  - r^{-1} \partial_r^2 \, r  \, \eqno (1) $$
is positive and Hermitian with respect to the usual scalar product
space in spherical coordinates
$$ \big\langle \phi_{{}_1} \, | \, \phi_{{}_2} \big\rangle
\equiv \int_\Omega \int_0^\infty \, \phi^*_{{}_1} \, \phi^{}_{{}_2}
\; r^2 \, dr \, d\Omega \, . \eqno (2) $$
The bounded eigenfunctions of  $ p_r^2 \, $ satisfying
$$ p_r^2 \phi = k^2 \phi \, \eqno (3) $$
are
$$ \phi_k = \sin ( k r ) / r \, . \eqno (4) $$
The singular functions $ \, [ \cos ( k r ) / r ] \, $ are
excluded as solutions of $ - \nabla^2 \phi = k^2 \phi \, $ because \hfil\break
$ \, - \nabla^2 \, [ \cos ( k r ) / r \, ] \, $ results in a $ \delta  $
function at the origin as
was pointed out by Dirac (p154 of [1]), so one imposes the boundary
condition [1]
$$ r \phi = 0 \qquad \hbox{for } \qquad r = 0 \, . \eqno (5) $$
The necessity of this boundary condition
arises from the singular nature of the spherical coordinate system at
the origin.

Various authors have pointed out that the operator
$$ \tilde{p}_r \equiv - i r^{-1} \partial_r r
= {1 \over 2} \, \big[ {{\bf r} \over r} \cdot ( - i \, {\bx \nabla} )
+ ( - i \, {\bx \nabla} ) \cdot {{\bf r} \over r} \big] \, $$
is not Hermitian despite its symmetric structure: the underlying reason being
that the radial integration within the scalar product (2) is over the
half axis instead of the whole axis.
Supposing for the moment that $ \tilde{p}_r \, $ is Hermitian, then there
exists the  unitary operator
$ \, [ \exp ( i a \tilde{p}_r ) ] \, $ generating the transformation
$ r \rightarrow \, r - a \, ,$
so the spectrum of $ r \, $ must extend from $ \, - \infty \, $ to
$ \, \infty \, $ which is clearly false as $ r \, $ is bounded below [2].
And the displacement  $ r \rightarrow \,
r - a \, $ is clearly unphysical as it would tear a hole at the origin.
For an extensive
discussion of the non-Hermiticity of $ \tilde{p}_r \, ,$
see [3].

For a positive Hermitian operator such as $ p_r^2 \, $ there exists a
unique square root Hermitian operator which we will call $ p_r^+ \, $
such that $ p_r^+ \, p_r^+  = p_r^2  \, $ and $ p_r^+ $
is positive.
To simplify matters we will eliminate the measure $ r^2 \, $
in the scalar product (1) by making the substitution
$$ \phi (r) = \chi (r) / r \, ,$$ then
the eigenvalue equation (3) is simply
$$ - \partial_r^2 \, \chi = k^2 \chi \, $$
and the scalar product space is
$ L^2 (0 , \infty ) \, :$
$$ \langle \chi_{{}_1} \, | \, \chi_{{}_2} \rangle_r
\equiv \int_0^\infty \, \chi^*_{{}_1} \, \chi^{}_{{}_2}
\; dr \,  \eqno (6) $$
with the boundary condition corresponding to (5)
$$  \chi (0) = 0 \, . \eqno (7) $$
Once we have found the square root operator of $ - \partial_r^2 \, $
which we will call $ z_r^+ \, ,$
which is Hermitian with respect to the space (6), then
$$ p_r^+ = ({1 \over r} \, z_r^+ \, r ) \, . \eqno (8) $$

We will show that the required operator $ z_r^+ \, $ is
$$ \eqalignno{
z_r^+ \,
&= {\cal H}_e  \,  \partial_r \,
= \, - \, \partial_r {\cal H}_o  \, & (9) \cr
  }  $$
where $ {\cal H}_e , \, {\cal H}_o \, $ are the Hilbert transforms of
even, odd functions defined by
$$ \eqalign{
{\cal H}_e \, f(r) &\equiv \, - \, { 2 r \over \pi } \,
\int_0^\infty { f(t) \over r^2 - t^2 } \, dt \, \cr
{\cal H}_o \, f(r) &\equiv \, { 2 \over \pi } \,
\int_0^\infty { t \, f(t) \over r^2 - t^2 } \, dt \, , \cr
  }  $$
the integral being a Cauchy principle value. It is known that
$ {\cal H}_e \, $ is the Fourier sine transform of the
Fourier cosine transform, while $ {\cal H}_o \, $ is
the Fourier cosine transform of the Fourier sine transform, i.e.
$$ \eqalign{
{\cal H}_e \, &\equiv {\cal F}_s {\cal F}_c \,  \cr
{\cal H}_o \, &\equiv {\cal F}_c {\cal F}_s \,   \cr
  } \eqno (10)  $$
where
$$ \eqalign{
{\cal F}_s \, f(r) &\equiv \, \sqrt{2 \over \pi} \,
\int_0^\infty  f(t) \, \sin (r t ) \, dt \, \cr
{\cal F}_c \, f(r) &\equiv \, \sqrt{2 \over \pi} \,
\int_0^\infty  f(t) \, \cos (r t ) \, dt \, . \cr
  }  $$
Noting the operator identities
$$ \eqalign{
{\cal F}_s \, \partial_r
&= \, - \, r \, {\cal F}_c \, \cr
{\cal F}_c \, \partial_x
&= \,+ \, r \, {\cal F}_s \, \cr  \cr
  } \eqno (11)  $$
then substituting (10) and (11) into (9) we obtain
$$ \eqalignno{
z_r^+ \, &\equiv \, {\cal H}_e  \,
\partial_r \,
= \, {\cal F}_s \, {\cal F}_c \, \partial_r \,
= \, {\cal F}_s \, r \, {\cal F}_s \, .
& (12) \cr
  }  $$
And
$$ \eqalignno{
( z_r^+ )^2 \,
&= \, ( \, {\cal F}_s \, r \, {\cal F}_s \, ) \,
( \, {\cal F}_s \, r \, {\cal F}_s \, ) \,
= \, {\cal F}_s \, r^2 \, {\cal F}_s \,
= \, - \, \partial_r^2 \, & (13) \cr
  }  $$
which demonstrates the required property $ ( z_r^+ )^2 \,
= \, - \, \partial_r^2 \, .$ For (12) we have used the identity
$ {\cal F}_s \, {\cal F}_s \, = \, 1 \,
= \, {\cal F}_c \, {\cal F}_c \, .$
The result (12) can also be obtained more directly from (9) using the identity
$ {\cal H}_e \, {\cal H}_o \, = {\cal H}_o \, {\cal H}_e \, = 1 \, .$

\noindent {\it Eigenfunctions of $ z_r^+ \, $ } \hfil\break
It can be verified that the
eigenfunction $ \, \sin ( k r ) \, $ of $  \, ( - \, \partial_r^2 ) \, $
is also an eigenfunction of $ z_r^+ \, ,$ as
$$ \eqalignno{
z_r^+ \,  \sin ( k r )  \,
&\equiv \, {\cal H}_e  \,  \partial_r \, \sin ( k r )  \,
= k \, {\cal H}_e  \, \cos ( k r )  \, = k  \,  \sin ( k r ) \, ,
\qquad \qquad k > 0 \, .\cr
  }  $$
Note that $ \cos ( k r ) \, $ is not an eigenfunction of $ z_r^+ \, ,$ in fact
$$ \eqalignno{
z_r^+ \, \cos ( \lambda r ) \,
&= - \, k \, {\cal H}_e  \, \sin ( k r )  \,
=  {2 k\over \pi} \, [ \sin ( k r ) \, \hbox{Ci } ( k r ) \,
- \, \cos ( k r ) \, \hbox{Si } ( k r ) \, ]
\, .  \cr
  }  $$

\noindent {\it $ z_r^+ \, $ is a positive operator  } \hfil\break
\noindent The fact that $ z_r^+ \, $ is a positive Hermitian operator
can be seen from the result (12) which is $  z_r^+ \,
= \, {\cal F}_s \, r \, {\cal F}_s \, .$ The operator $ {\cal F}_s \, $
is Hermitian with respect to the
$ L^2 (0 , \infty ) \, $ space (6), which can be shown by writing out
$ \big\langle \, \psi \, \big| \, {\cal F}_s \, \chi \, \big\rangle \, .$
then changing the order of integration, i.e.
$$ \eqalignno{
\big\langle \, \psi \, \big| \, {\cal F}_s \, \chi \, \big\rangle \,
&\equiv \int_0^\infty \, \psi ^* (r) \,
\left[ \int_0^\infty \, \sin ( r t ) \, \chi (t) \, dt \, \right] \, dr \,  \cr
&= \int_0^\infty \, \chi (t) \,
\left[ \int_0^\infty \, \psi ^* (r) \, \sin ( r t ) \,
dr \, \right] \, dt \, = \big\langle \, {\cal F}_s \, \psi \,
\big| \, \chi \, \big\rangle \, . & (14)
  }  $$
Then using (14)
$$ \eqalignno{
\big\langle \, \chi \, \big| \, z_r^+ \, \chi \, \big\rangle \,
&= \big\langle \, \chi \, \big| \, {\cal F}_s \, r \, {\cal F}_s \, \chi \,
\big\rangle \,
= \big\langle \, \sqrt{r} \, {\cal F}_s \, \chi \, \big| \,
\sqrt{r} \, {\cal F}_s \, \chi \, \big\rangle \, \ge \, 0 \, . & (15) \cr
  }  $$

\noindent {\it The inverse of $ z_r^+ \, $ } \hfil\break
\noindent Every positive operator has an inverse which is also positive.
From (12) we see that
$$ \eqalignno{
( z_r^+ )^{-1} \, &= \, {\cal F}_s \, {1\over r} \, {\cal F}_s \, .
& (16) \cr
  }  $$
which is clearly positive by similar reasoning to (15). Writing out (16)
$$ \eqalignno{
( z_r^+ )^{-1} \, \chi (r) &\equiv \int_0^\infty \, dt \, \sin ( r t ) \,
{1\over t} \,
\int_0^\infty \, \sin ( u t ) \, \chi (u) \; du  \; \cr
\noalign{\noindent
and carrying out the integration with respect to $ t \, $ we obtain }
( z_r^+ )^{-1} \, \chi (r) &= {1 \over 2 } \,
\int_0^\infty \, \log \left| { r + u \over r - u } \right|  \, \chi (u) \; du
\, . \cr
  }  $$

\noindent {\it $ ( z_r^+ )^{-1} \, $ in terms of the
fractional integral operators } \hfil\break
\noindent Erdelyi [4] and Kober [5] established relations between
the Hankel transforms
(of which $ {\cal F}_s , \, {\cal F}_c \, $ are special cases) and the
so-called fractional integral operators (sometimes called Erdelyi-Kober
operators after their originators). In particular
$$ \eqalignno{
( z_r^+ )^{-1} \, &= \, {\cal F}_s \, {1\over r} \, {\cal F}_s \,
= r \, {\cal K} \, {\cal I} \,
& (17) \cr
  }  $$
where the Erdelyi-Kober operators $ {\cal I} , \, {\cal K} \, $ are defined by
$$ \eqalign{
&{\cal I} \, f( r ) \equiv {2 \over \sqrt{\pi}} \,
\int_0^1 \, u \; ( 1 - u^2 )^{- 1/2 }  f( u r ) \; du \cr
&{\cal K} \, f( r ) \equiv {2 \over \sqrt{\pi}} \,
\int_1^\infty  ( u^2 - 1 )^{- 1/2 } \;  f( u r ) \; du \, . \cr
 } \eqno (18) $$
Due to the Hermitian conjugate relations [4]
$$ ( r \, {\cal K} )^{\dag} = r \, {\cal I} \,
\qquad \qquad  ( r \, {\cal I} )^{\dag}
= r \, {\cal K} \,  $$
then
$$ \eqalignno{
\big\langle \, \psi \, \big| \, ( z_r^+ )^{-1} \chi \, \big\rangle \,
&= \big\langle \, \psi \, \big| \, r \, {\cal K} \, {\cal I} \, \chi \,
\big\rangle \,
= \big\langle \, r \, {\cal I} \, \psi \, \big| \,
{\cal I} \, \chi \, \big\rangle \, & (19)  \cr
  }  $$
which is another way of showing that $ ( z_r^+ )^{-1} \, $ is positive.
If we insert the eigenfunctions of $ z_r^+ \, $ into (19),
with $ \chi = \sin ( k r ) , \,
\psi = \sin ( k' r ) \, ,$ and as
$$ {\cal I} \, \sin ( k r ) =  \sqrt{ \pi } \, J_1 ( k r ) \, $$
we arrive at
$$ \eqalignno{
\big\langle \, \sin ( k' r ) \big| \,
( z_r^+ )^{-1} \sin ( k r )  \, \big\rangle \,
&= {\pi } \, \big\langle \, r \, J_1 ( k' r ) \, \big| \,
J_1 ( k r ) \, \big\rangle \,
= {\pi \over k } \, \delta (k - k' )  \cr
  }  $$
which is an interesting relation between the orthogonality of sine functions
and Bessel functions.
The operators $ {\cal I} , \, {\cal K} \, $ are invertible, for a
succinct introduction to these operators see chapter 2 of [6].
Rooney [7] used the operator identities
$$ \eqalign{
{\cal H}_e &= {\cal K}^{ -1} \, r \, {\cal I} \, {1 \over r}  \qquad \qquad
{\cal H}_o =  r \, {\cal I}^{ -1} \, {1 \over r} \, {\cal K} \, \cr
 }  $$
to investigate the ranges of $ {\cal H}_e , \, {\cal H}_o \, .$

\noindent {\it Conclusion } \hfil\break
While there has been discussion in the literature of the $ \tilde{p}_r \, $
operator and its non-Hermiticity, the $ p_r^+ \, $ operator
$$  p_r^+ = {1 \over r} \, z_r^+ \, r \,
= {1 \over r} \, {\cal H}_e  \,  \partial_r r \, $$
appears to be little known.
Unfortunately the $ p_r^+ \, $ operator has no simple commutation relation
with $ r \, .$
A possible application of this operator is in light-cone quantum theory,
where the Hamiltonian contains an inverse radial momentum operator - see
[2] which addresses this problem and the references therein, or the original
work by Dirac [8].

\beginsection References

\item{[1]} Dirac P. A. M., {\it The Principles of Quantum Mechanics}
(OUP, Oxford, 1947) 3rd ed.
\item{[2]} Levin O. and Peres A., ``Some oddities of lightcone dynamics,''
J. Phys. A {\bf 27}, L143-145 (1994)
\item{[3]} Liboff R. L. et al, ``On the radial momentum operator,'' Am. J. Phys.
{\bf 41}, 976-980 (1973)
\item{[4]} Erdelyi A., ``On fractional integration and its application to the
theory of Hankel transforms,''
Quart. J. of Math. Oxford {\bf 11}, 293-303 (1940)
\item{[5]} Kober H., ``On fractional integrals and derivatives,''
Quart. J. of Math. Oxford {\bf 11}, 193-211 (1940)
\item{[6]} Sneddon I. N., {\it Mixed boundary problems in potential theory}
(North Holland, Amsterdam, 1966)
\item{[7]} Rooney P. G., ``On the ranges of certain fractional integrals,''
Can. J. Math. {\bf 24}, 1198-1216 (1972)
\item{[8]} Dirac P. A. M.,  ``Forms of relativistic dynamics,'' Rev. Mod. Phys.
{\bf 21}, 392-399 (1949)

\bye